# Controlling the electrical and magnetic ground states by doping in the complete phase diagram of titanate Eu$_{1-x}$La$_x$TiO$_3$ thin films

Hyungki Shin,[1,2,*] Chong Liu,[1,2] Fengmiao Li,[1,2] Ronny Sutarto,[3] Bruce A. Davidson,[1,2] Ke Zou[1,2,*]

[1] Department of Physics and Astronomy, University of British Columbia, Vancouver, BC V6T 1Z1, Canada
[2] Quantum Matter Institute, University of British Columbia, Vancouver, BC V6T 1Z4, Canada
[3] Canadian Light Source, Saskatoon, SK S7N 2V3, Canada

EuTiO$_3$, a band insulator, and LaTiO$_3$, a Mott insulator, are both antiferromagnetic with transition temperatures ~ 5.5 K and ~ 160 K, respectively. Here, we report the synthesis of Eu$_{1-x}$La$_x$TiO$_3$ thin films with $x = 0$ to 1 by oxide molecular beam epitaxy. The films in the full range have high crystalline quality and show no phase segregation, allowing us carry out transport measurements to study their electrical and magnetic properties. From $x = 0.03$ to 0.95, Eu$_{1-x}$La$_x$TiO$_3$ films show conduction by electrons as charge carriers, with differences in carrier densities and mobilities, contrary to the insulating nature of pure EuTiO$_3$ and LaTiO$_3$. Following a rich phase diagram, the magnetic ground states of the films vary with increasing La-doping level, changing Eu$_{1-x}$La$_x$TiO$_3$ from an antiferromagnetic insulator to an antiferromagnetic metal, a ferromagnetic metal, a paramagnetic metal, and back to an antiferromagnetic insulator. These emergent properties reflect the evolutions of the band structure, mainly at the Ti $t_{2g}$ bands near the Fermi level, when Eu$^{2+}$ are gradually replaced by La$^{3+}$. This work sheds light on this method for designing the electrical and magnetic properties in strongly-correlated oxides and completes the phase diagram of the titanate Eu$_{1-x}$La$_x$TiO$_3$.

## I. INTRODUCTION

The search for new materials with novel functional properties has driven the synthesis of complex oxides in the forms of thin-films, heterostructures, and superlattices over the last few decades. In transition-metal $ABO_3$ perovskites, A–site substitution allows control of the B–site valence as well as induced ordering of the $BO_6$ octahedral rotations via structural distortions due to changes in the tolerance factor [1]. Both effects together with Jahn-Teller distortions [2] and charge and bond disproportionation [3] among others can lead to strongly-correlated systems that show emergent properties such as metal-insulator transitions, magnetism, or superconductivity. The general question of the bandwidth- and band filling-dependent properties of solid solutions between perovskite end members has been addressed [4] and continues to be of extreme interest [5]. For example, $Sr^{2+}$ substitution for $Nd^{3+}$ in the oxygen-deficient perovskite $Nd_{1-x}Sr_xNiO_2$ converts an insulator to a superconductor [6], and substitution of $Sr^{2+}$ for $La^{3+}$ in $LaTiO_3$ changes the antiferromagnetic (AF) Mott insulator into paramagnetic metallic $La_{1-x}Sr_xTiO_3$ (SLTO) [7–10]. Similarly, the phase diagrams for single crystals of $La_{1-x}Sr_xMnO_3$ [11] and thin films of $La_{1-x}Sr_xMnO_3$ [12] and $La_{1-x}Sr_xFeO_3$ [13] display a wide range of magnetic and electronic phases.

Perovskites containing Eu at the A–site, such as $EuTiO_3$ (ETO) and $EuNiO_3$, show distinctive magnetic properties because of the partially-filled $Eu^{2+}$ 4$f$ orbitals. ETO has a similar structure as the well-studied $SrTiO_3$ with identical 3.905 Å lattice parameter, space group Pm$\bar{3}$m, and a band gap of ~1 eV [14]. The projected bands near Fermi level have mainly hybridized Ti 3$d$ – O 2$p$ character, with Eu 4$f$ bands that lie ~1 eV lower in the valence bands [Fig. 1(a)]. ETO undergoes a structural phase transition below room temperature [15] showing coupling between magnetic and dielectric properties [16] and a G-type AF ordering below Néel temperature $T_N$ = 5.5 K with 7 Bohr magnetons ($\mu_B$) on each $Eu^{2+}$ site due to the unpaired and localized 4$f^{\,7}$ electrons. More recent studies on weakly-doped metallic ETO thin films show unusual anomalous Hall effect (AHE), an indication of different long-range order that changes with doping [17,18]. Magnetic phase transitions due to substitution and strain have also been reported [19–21].

Perovskite $LaTiO_3$ (LTO), a strongly correlated Mott insulator with bandgap 0.2 eV from the splitting of the Ti 3$d$ orbitals [22], is orthorhombic with a pseudocubic lattice constant of 3.97 Å, and becomes AF with G-type ordering below $T_N$ =150-160 K with 0.5 $\mu_B$ per Ti site [2,23,24]. It also generates an interfacial two-dimensional electron gas (2DEG) when deposited on $SrTiO_3$ [25,26] and $KTaO_3$ [27] substrates due to the structural distortion or polar discontinuity. Previously, the substitution of $La^{3+}$ by $Sr^{2+}$ in LTO induces a metal-insulator transition with hole-carrier type for <10% substitution level, and electron-carrier type at higher levels [23,28]. In contrast to Sr doping, Eu doping is expected to increase electron correlations in Ti 3$d$ bands due to interactions between Eu 4$f$ and lower Hubbard band (LHB) near the Fermi level [7]. This will simultaneously introduce additional degree of freedom in the spin-polarized Eu 4$f$ state. The full range of intermixing between ETO and LTO has not been studied in theory and experiment and it still is an open question how the Eu occupied 4$f$ state close to the Fermi level interacts with the Ti 3$d$ state [Fig. 1(a)].

Here, we report the investigation of solid solution of ETO and LTO (Eu$_{1-x}$La$_x$TiO$_3$, ELTO) over the range 0< $x$ <1 corresponding to the evolution from band insulator to Mott insulator. Because of the similar lattice constant and structure, molecular beam epitaxy (MBE) is able to create high-quality epitaxial films with A-site intermixing of La$^{3+}$ and Eu$^{2+}$ atoms across the entire phase diagram. All films are metallic except for within a few percent of the end members, allowing the electronic and magnetic ground states of ultrathin films (10 unit cells) to be characterized by magneto transport measurements. Hall measurements show only electron carriers for all doping levels, an intriguing progression from AF → Ferromagnetic (F) → Paramagnetic (P) → AF as $x$ = 0→1, monotonically disappearing AHE at $x$ = 0.5, and a mobility collapse above $x$ = 0.5 due to the onset of correlations.

## II. SAMPLE FABRICATION

All films were grown on insulating (LaAlO$_3$)$_{0.3}$(Sr$_2$AlTaO$_6$)$_{0.7}$ (LSAT) (001) substrates (CrysTec GmbH, Berlin) to insure insulating behavior in pure EuTiO$_3$ and LaTiO$_3$ and minimize oxygen vacancies at the relatively low oxygen partial pressure during growth. LSAT remains highly insulating (>1 GΩ) after thermal cycles in UHV, and no 2DEG has been reported at interfaces of LSAT with other oxides. The lattice constant of LSAT is 3.868 Å, smaller than EuTiO$_3$ (3.905 Å) and LaTiO$_3$ (3.97 Å), and the resulting epitaxial strain is compressive in all films. Calculations suggest this strain is not enough to change the magnetic properties in pure EuTiO$_3$ films [21].

The samples were grown in an oxide molecular beam epitaxial (MBE) system (Veeco GenXplor) with a base pressure < 1x10$^{-10}$ torr. Europium, titanium, and lanthanum are thermally evaporated from the effusion cells under molecular oxygen or plasma oxygen. The fluxes of these elements were calibrated by the quartz crystal microbalance (QCM) before growth and later confirmed by *in situ* reflection high-energy electron diffraction (RHEED) oscillations. The level of chemical substitution is determined by the ratio of QCM fluxes. The films were grown at substrate temperature 680°C in a partial pressure of molecular oxygen of 3x10$^{-8}$ Torr. Since EuTiO$_3$ and LaTiO$_3$ are easily oxidized in ambient atmosphere once removed from the MBE system, all samples are capped *in situ* with 5 nm of Ti + amorphous TiO$_x$ at room temperature for *ex situ* measurements. The capping layer is designed to protect the films underneath, without making any contribution to the transport. It was checked the capping layer and confirmed clear insulating behavior in Ti/TiO$x$ film on LSAT substrate and pure ETO (x= 0) and LTO (x=1) films.

Devices were fabricated in the van der Pauw geometry for electrical transport measurements. To measure the underlying ELTO thin films, the TiO$x$ capping layer was scratched by diamond knife and indium was soldered immediately between the gold wire and the film to make better contact. Indium soldered pads with gold wires were used as electrodes for longitudinal ($R_{xx}$) and transverse ($R_{xy}$) measurement in physical property measurement system (PPMS, Quantum Design).

## III. EXPERIMENTAL RESULTS

### A. Surface morphology

Film growth was monitored by *in situ* RHEED, and characterized by *ex situ* Atomic force microscope (AFM) and X-ray diffraction (XRD) measurements after the growth. They confirm the high quality of the films and yield the lattice constants for substitution over the full range between $x = 0$ to 1. Figure 1(b) shows clear intensity oscillations from RHEED during the $x = 0.5$ growth, and other films show similar behavior. The oscillation period confirms the expected 10 uc thickness for all samples, as expected from the QCM fluxes. Figure 1(c)-(e) show the RHEED patterns of $x = 0, 0.5$, and 1 after growth, respectively. These *in situ* RHEED results indicate a smooth surface with no island growth for all films.

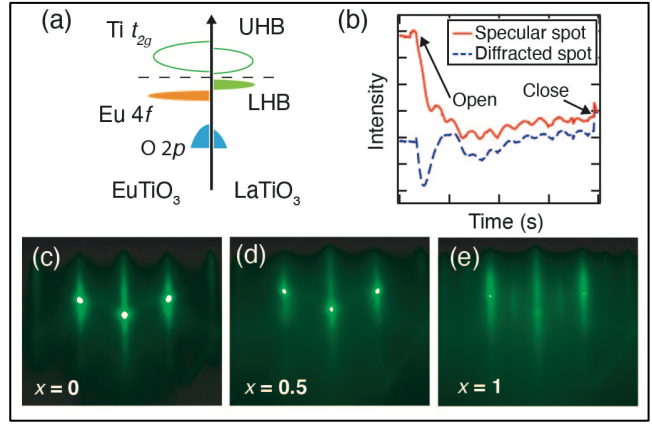

FIG. 1. (a) Schematic band structures of $EuTiO_3$ ($x = 0$) and $LaTiO_3$ ($x = 1$), where UHB and LHB are defined as upper and lower Hubbard band, respectively. (b) Reflection high-energy electron diffraction (RHEED) oscillations during the growth showing the layer-by-layer growth of atomically smooth films. (c-d) RHEED images of 10 unit cell (uc) $EuTiO_3$, $Eu_{0.5}La_{0.5}TiO_3$, and $LaTiO_3$ films, respectively.

Figure 2(a) and (b) show an AFM image of an uncapped ETO ($x = 0$) film, showing atomically flat terraces separated by single unit cell steps. XRD scans in Fig. 2(c) show sharp diffraction peaks on all films with clear Laue fringes, indicating high crystalline quality. The diffraction peak of ELTO gradually moves to lower angles for increasing $x$, indicating an increasing out-of-plane lattice parameter and correspondingly higher strain. We extract the films' in-plane lattice parameters from the spacing between RHEED diffraction spots, and out-of-plane parameters from XRD scans [Fig. 2(d)]. The in-plane parameter is identical to the substrate within the error bars, indicating all ELTO films are fully strained to the $(LaAlO_3)_{0.3}(Sr_2AlTaO_6)_{0.7}$ (LSAT) substrate. The strain $\varepsilon = (a_\parallel - a_o)/a_o$ can be estimated between –0.95% (ETO) and -2.57% (LTO), where $a_0$ is the lattice parameter of bulk single-crystal and $a_\parallel$ is the measured film in-plane lattice constant.

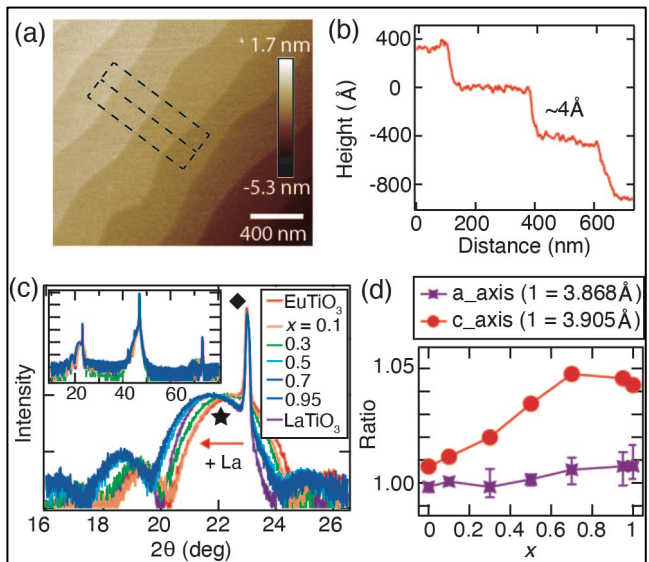

FIG. 2. (a) AFM image of an uncapped 10 uc ETO film surface. (b) Line profile of the dashed region in (a) showing step and terrace structure, with ~ 4 Å = 1 ETO unit cell. (c) Symmetric XRD scans of 10 uc $Eu_{1-x}La_xTiO_3$ (ELTO) films around (001) peak, ◆ at $2\theta$ = 22.99° and ★ near $2\theta \sim 22°$ are peaks from LSAT substrate and ELTO films, respectively. (d) The normalized ratio of film in- and out-of-plane lattice parameters $a$ and $c$ as a function of substitution level $x$ (Purple: $a_{ELTO}/a_{LSAT}$; Red: $c_{ELTO}/a_{LSAT}$.). The fitting and error bars of $a$-axis are obtained from the line spacing and its standard deviation between diffracted spots in the RHEED image. The fitting and the error bars of $c$-axis are obtained from XRD spectra with the error bars smaller than the symbols.

### B. Electrical transport measurement

We characterize the films' electrical and magnetic properties via electrical and magneto transport measurements [Figs. 3 and 4]. Undoped ETO and LTO films both show high resistivity ($\rho_{RT}$>10 mΩcm) and insulating behavior in transport measurement below 300 K [inset of Fig. 3(a)]. Consequently, we can rule out bulk doping or significant amounts of oxygen vacancies in the films [26,29–32].

Metallicity should be introduced for 0< $x$ <1 when La behaves as an electron donor ($La^{3+}$) in ETO, and Eu behaves as a hole donor ($Eu^{2+}$) in LTO. Figure 3(a) shows the change of resistivity ($\rho - \rho_{RT}$) as a function of $T^2$ and $T$ for 0< $x$ <1 and additional $\rho$ data at 300 K and 3 K are shown in Fig. 3(b). As seen in the inset, at low substitution levels ($x$ = 0.01 and 0.97) the films show insulating behavior ($dR_{xx}/dT < 0$) similar to LTO and ETO films. This is in agreement with previous studies of insulating behavior in pure and lightly-doped ETO and LTO [18,28,30,33,34].

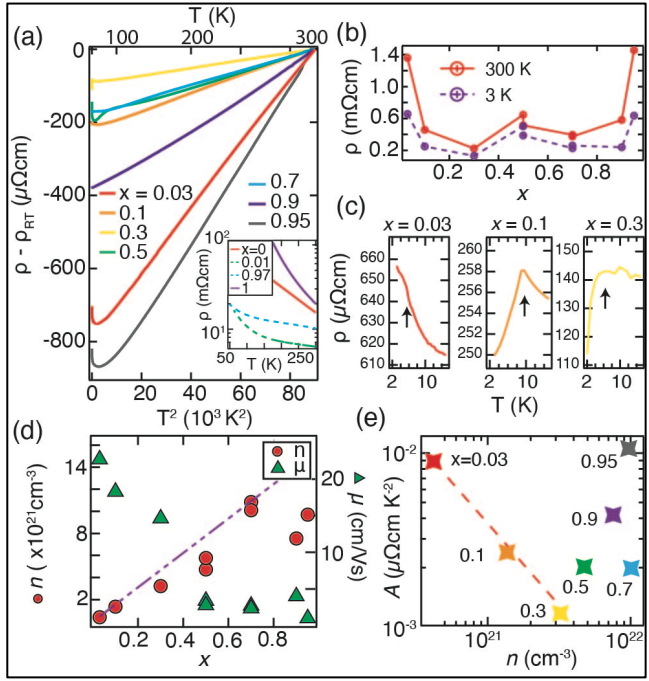

FIG. 3. (a) The resistivity vs $T^2$ and $T$ for different substitution level $x$, where $\rho_{RT}$ is resistivity at 300 K. Inset: insulating behavior of $x = 0, 0.01, 0.97,$ and 1. (b) The resistivity vs $x$ at 3 K and 300 K. (c) Kinks present in $\rho$ vs $T$ (d) Charge carrier density $n$ (left axis, red circle) and mobility $\mu$ (right axis, green triangle) versus $x$ at 3 K. $n$ are extracted from the Hall measurement under perpendicular B-field in ± 7 T at 3 K, and are constant between 3 and 20 K. The purple dashed line indicates 100% activated carrier density from the assumption that each La substitution supplies one electron. (e) Coefficient of the $AT^2$ resistivity versus carrier density $n$. The dashed line is a guide to the eye for $A \propto n^{-1}$ from the three-band model [31].

We observe signs in the $R(T)$ curves that correspond to transitions of the magnetic states. All metallic samples show a slight increase of resistivity below 30 K. Furthermore, the resistivities of $x = 0.03, 0.1,$ and 0.3 films show kinks at 5.5 K, 8.5 K, and 6.3 K, respectively, corresponding to the $T_N$ or Curie temperature ($T_C$) determined by their magnetic phase [Fig. 3(c)]. For $x = 0.1$ and 0.3, the resistivity quickly decreases below the transition temperatures. Increasing resistivity between 30 K and the low-temperature kink in resistivity vs $T$ has been reported previously in studies of magnetic films and has been ascribed to strong interactions between itinerant electrons in Ti 3$d$ bands and localized spins in Eu 4$f$ [19]. Below the kink temperature, spins are aligned AF or F and the resistivity decreases. No kink was observed above $x = 0.5$, suggesting the lack of (or weak) magnetic ordering.

We plot carrier density and mobility at 3 K from Hall measurements [Fig. 3(d)]. The measured $R_{xy}$ are linearly fitted above $3T$ to exclude the AHE near zero-field as a form of non-linear slope shown by others [17,18,33]. All metallic films show a negative slope in $R_{xy}$ vs $B$, indicating electron-type carriers. Close to the ETO side, this carrier type is expected since $La^{3+}$ donates an extra electron to $Eu^{2+}Ti^{4+}O_3$. However, the presence of electron carrier type even close to LTO is curious because $Eu^{2+}$ is expected to donate holes in the $La^{3+}Ti^{3+}O_3$ system. Similar results have been reported in studies of Sr-doped LTO, for which Sr substitution >10% has electron-type carriers, not hole-type. It is possible that strain effect modifies the orbital configuration of Mott insulator LTO leading to a change in the energy bandwidth [28,35] or that the Fermi level passes through the strongly-localized LHB [Fig. 1(a)], and the localized holes show insulating behavior, such as $x = 0.97$ in our case.

The ratio of activated carrier density to the substitution level is close to one for all samples [Fig. 3(d)]. Deviation from 100% activation may be influenced by the crystalline quality of films or be determined by the band structure. A similar activated carrier level has been reported in other studies with different substitutional elements in perovskites [28,33,36] Detailed theoretical calculations for ELTO are required to explain these doping density results together with the band structure evolution.

The mobilities at 3 K from Hall measurements gradually decrease from 22 cm$^2$/Vs at $x = 0.03$ to 1 cm$^2$/Vs at $x = 0.95$. Interestingly, the largest decrease of mobility is seen near $x = 0.5$ and higher, possibly due to the increasing number of intermixing atoms that acting as "charged impurities" in transport, although significant structural defect is not shown in RHEED and XRD. The general trend of reduced mobility with increasing $x$ may also be explained by the different ionic radii of Eu$^{2+}$ and La$^{3+}$ that induce structural distortions resulting in variation of the Ti-O-Ti bond angle, from near 180° in bulk ETO to 157° in bulk LTO [31,37,38]. This distortion leads to a reduction of the hopping integral, possibly decreasing mobility as a result.

In addition, La substitution in ETO changes the Ti valence from tetravalent (Ti$^{4+}$) to trivalent (Ti$^{3+}$) with a corresponding increase of the Coulomb repulsion in the system, indicative of the onset of a strong electron correlations [4,39]. As the strongly-correlated system is gradually formed from pure ETO, the effective mass can be enhanced [7,38] so that mobility would decrease, as seen from the Drude relation $\mu = e\tau/m^*$ where $\tau$ is the momentum relaxation time and $m^*$ the effective mass. Photoemission studies of SLTO supports the interpretation that the drastic decrease of mobility could be induced in the strongly-correlated system by the formation of new Hubbard band beginning near Sr$_{0.5}$La$_{0.5}$TiO$_3$ [40]. To clarify this, band structure studies near $x = 0.5$ are needed since ELTO has a more complicated electronic structure than SLTO due to polarized Eu 4$f$ spins near the Fermi level and interaction between Eu 4$f$ and Ti 3$d$ bands.

For $0.03 < x < 0.95$, all films show metallic behavior (d$R_{xx}$/d$T > 0$) well-characterized by $\varrho = AT^2+B$ above 10 K [Fig. 3(a)] suggesting that electron-electron interactions are the main source of scattering. The $T^2$ behavior confirms that oxygen deficiency is not the primarily reason for metallicity, as oxygen deficiency in ETO$_{3-\delta}$ shows a resistivity deviating from $T^2$ where α in $T^\alpha$ varies with carrier density [31]. Figure 3(e) show the relation between the coefficient A from fitting the transport data in Fig. 3(a) and carrier density n, and the dashed line indicates the power law model adapted from Ref. [31] based on the three-band system observed in ETO. The points from $x = 0.03$ to 0.3 clearly follow this power law model implying that their transport behavior can be explained by three-band model. However, the films after $x = 0.5$ significantly deviate from this model while their carrier densities are still in the range of three-band system, suggesting a more complicated band structure above $x = 0.5$. Likewise, the systematic enhancement of A after $x =$ 0.5, suggesting stronger on-site Coulomb interaction, can be the signal of approaching to Mott-Hubbard band system [4,7].

**C. Magneto transport measurement**

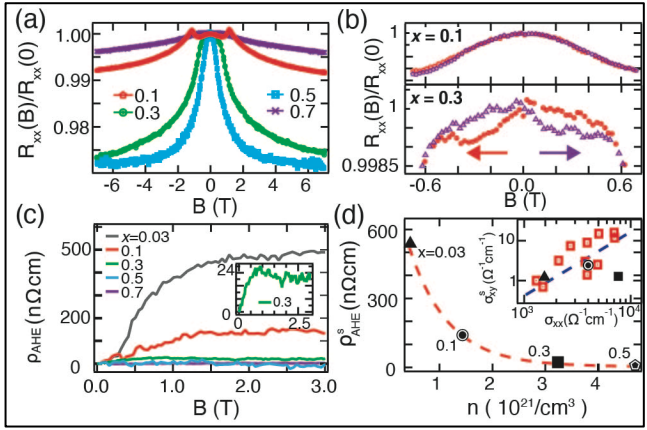

FIG. 4 (a) Magnetoresistance $R_{xx}$ vs $B$ at 3 $K$ measured under perpendicular B-field swept from 7 $T$ to -7 $T$ and back at 3 $K$. (b) Zoomed in data around zero magnetic field, ± 0.6 $T$ for $x$ = 0.1 and 0.3. (c) Anomalous Hall effect (AHE) $\rho_{AHE}$ extracted from conventional Hall measurement ($\rho_{xy} - \rho_{Hall}$). $\rho_{AHE}$ is saturated after 3 $T$ in all samples. Inset: zoomed in data for $x$ = 0.3 near zero magnetic field. (d) Saturated anomalous Hall resistivity ($\rho_{AHE}^s$) as a function of charge carrier density. The dashed line is a guide to eye. Inset: The relation between anomalous Hall conductivity ($\sigma_{AHE}^s$) and longitudinal conductivity ($\sigma_{xx}$). The blue dashed line shows the relation $\sigma_{AHE}^s \propto \sigma_{xx}^{1.6}$ indicating the Berry's phase mechanism for AHE. The data with symbols □ is taken from Ref. [17].

We note, due to the thickness (~4 nm) of these films, SQUID magnetometer becomes inadequate to detect such low magnetic signals from these titanates. One indication is the lack of such measurements in any LTO films. Hysteresis was checked in the magnetoresistance (MR) $R_{xx}$ for $x$ = 0.1–0.7 [Fig. 4(a) and 4(b)]. The magneto transport measurements reveal the magnetic ground states and transition temperatures in these films. All films show negative MR without hysteresis except for $x$ = 0.3. It should be mentioned that no hysteresis was observed in Hall measurements ($R_{xy}$) in the raw data of any film but only in $R_{xx}$ for $x$ = 0.3, possibly due to small magnetization in F phase, or the easy axis perpendicular to the applied field.

Figure 4(c) shows AHE in several samples extracted from Hall measurement. $\rho_{AHE}$ are extracted after antisymmetrizing raw data of $R_{xy}$ ($R_{xy} = [R_{xy\_raw}(B)-R_{xy\_raw}(-B)]/2$) and subtracting the conventional Hall effect component (linear term, $\rho_{AHE} = \rho_{xy} - \rho_{Hall}$). First, $x$ = 0.03 shows the highest value of AHE saturating above 2 $T$ [Fig. 4(c)], indicating this film is AF and conducting [17,18]. The film $x$ = 0.1 shows distinctive peaks near ±1 $T$ in Fig. 4(a) and no hysteresis [Fig. 4(b)], as well as AHE [Fig. 4(c)] much smaller than $x$ = 0.03. These results without hysteresis indicate $x$ = 0.1 is also AF. The peaks near ±1 $T$ may result from spin scattering due to the AF to canted AF transition with increasing applied field. The peak and decline of MR has been shown in a previous study of transport in AF triggered by spin scattering [41].

The $x = 0.3$ film [Fig. 4(b) and 4(c)] shows small hysteresis in $R_{xx}$ vs $B$, and AHE, and its magnitude is similar to that seen in Sm-doped ETO [20]. Thus, we conclude that $x = 0.3$ is F and an AF-F transition occurs between $x = 0.1$ and 0.3. Between $x = 0.5$ and 0.7, interestingly, there is a significant change in Fig. 4(a), even both cases do not show the kink in Fig. 3(a) and AHE in Fig. 4(c). It is likely that $x = 0.5$ has a weak or canted F state, since $x = 0.3$ and 0.5 show similar behavior and magnitude of $R_{xx}$ vs $B$ and $x = 0.5$ shows a temperature dependence only below 7 K. The kink in Fig. 3(a) and hysteresis in MR are not observable for $x = 0.5$ likely because its net magnetization is below the limit detectable by MR measurements.

Figure 4(d) shows the saturated AHE resistivity ($\rho_{AHE}^S$) as a function of charge carrier density. AHE is extracted from the saturated region of $\rho_{AHE}$ above 3 $T$. Each symbol indicates the substitution level from $x = 0.03$ to 0.5, and clearly shows the AHE exponentially decays to zero as a function of carrier density after $x = 0.5$. This behavior and its magnitude are in agreement with previous studies in higher carrier density region. However, we did not observe any sign change of AHE, which has only been observed at low density ($<3\times10^{20}$ cm$^{-3}$) where Fermi energy is near the band-crossing point [17,18,32].

From the inset in Fig. 4(d), the mechanism of AHE can be investigated from the relation between $\sigma_{xx}$ and $\sigma_{xy}$, where $\sigma_{xx} = \rho_{xx}(B=0)/[\rho_{xx}^2(B=0) + (\rho_{AHE}^S)^2]$ and $\sigma_{xy} = \rho_{AHE}/[\rho_{xx}^2(B=0)+(\rho_{AHE}^S)^2]$. The power $\alpha$ in $\sigma_{xy} \sim \sigma_{xx}^\alpha$ enables determination of the mechanism among three possibilities: intrinsic deflection from Berry's phase curvature, side jump, and skew scattering [42]. $x = 0.03$ (▲) and 0.1 (◉) data closely follow the trend of $\alpha = 1.6$ indicating that AHE at these doping levels is induced by Berry curvature with disorder. However, the case of $x = 0.3$ deviates significantly from the line of $\alpha = 1.6$ meaning that the mechanism of AHE could not be the same. This could result from a highly disordered structure due to the high concentration of La substitution or magnetic phase transition to F state, or the solid substitution of La on Eu sites that can develop new band structure. To confirm the source of AHE, more studies in the F magnetic state are needed.

## IV. DISCUSSION AND CONCLUSION

The phase diagram as a function of substitution level $x$ is summarized in Fig. 5, with data from references. Each colored region corresponds to specific magnetic and electrical characteristics: antiferromagnetic insulator (gray-AFI), antiferromagnetic metal (blue-AFM), ferromagnetic metal (red-FM), and paramagnetic metal (yellow-PM). Close to ETO, $x = 0.03$ to $0.1$ films show AFM behavior with $T_N = 5.5$ K and 8.5 K, respectively. Between $x = 0.1$ to $0.3$ the films become FM with $T_C = 6.5$ K. With increasing doping, the film loses its magnetic property to canted/weak FM ($x = 0.5$) and then PM phases with slightly increased resistivity.

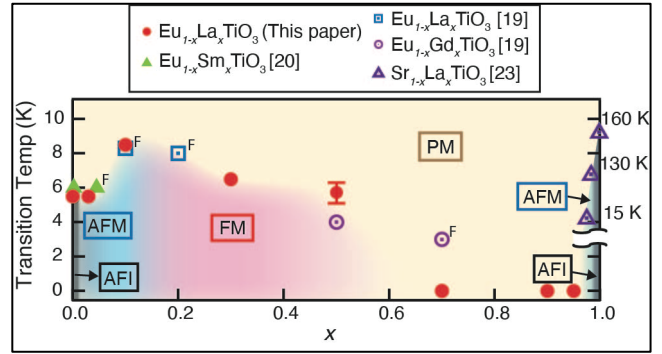

FIG. 5. Phase diagram of ELTO for its metallic and magnetic phase. Each colored region indicates different metallic/magnetic phase. AFI, AFM, FM, and PM represent antiferromagnetic insulator, antiferromagnetic metal, ferromagnetic metal, and paramagnetic metal, respectively. The symbol F indicates published FM phase data that fall outside the FM region reported here. Filled symbols and empty symbols represent the result from thin films and single crystal, respectively. The $T_C$ with error bar in $x = 0.5$ is extracted from the change of magnetoresistance in several temperatures.

Interestingly, the transition temperatures $T_C$ or $T_N$ and magnetic states vary depending on trivalent dopant element and the crystalline form, whether thin film or single crystal. For instance, two samples in $x = 0.1$ shown here have different magnetic states, AFM in ours and FM in single crystals from Ref. [19]. Also, Sm-doped ETO has a fixed transition temperature at low doping, even with different magnetic AF and F states [20]. There are several possible reasons for these variations: level of activated carrier density, whether substituted element could sustain the Eu spin network, and whether the unit cell is strained or freely relaxed. All these factors can affect spin exchange interactions, which will be discussed in the following section.

Mechanisms for these magnetic transitions have been discussed in previous studies [14,19,43] in which itinerant electrons from substituted donors in Ti $t_{2g}$ bands mediate indirect exchange interactions between localized spins in Eu $4f$, known as the Ruderman-Kittel-Kasuya-Yosida (RKKY) interaction; this effect can be enhanced by interaction through both Ti $3d$-Eu $4f$ and Ti $3d$-Eu $5d$ channels. This theory can readily explain AFI-AFM and AFM-FM transitions in our system as well as the variation of transition temperature, since it is expected more electrons (in the range $x < 0.5$) can mediate spin interactions more strongly. However, high concentration of La, such as in $x > 0.5$, can break the spin networks between Eu due to the different ionic radius and electron configuration between La and Eu, unlike the cases of Sm and Gd substitutions where $4f^7$ orbitals are present. In this scenario, the spin interaction vanishes after a certain level of La substitution resulting in the loss of magnetic properties to PM. Furthermore, filling electrons into $t_{2g}$ bands increase the on-site Coulomb interaction leading to the enhancement of effective mass, possibly related to the reduction of mobility in Fig. 2, further reducing spin interactions and decreasing the transition temperature.

In addition to this itinerant-electron-mediated mechanism, compressive strain can contribute to the magnetic phase transition to FM phase around $x = 0.3$. Theoretical predictions and experimental results show that compressive strain above 1.2% can induce a magnetic transition of pure ETO films from AFM to FM due to the spin-phonon (lattice) interaction [21,44,45] As the strain at $x = 0.3$ is 1.44%, beyond the value suggested by theory and experiment, compressive strain from La substitution can be another reason for a magnetic phase transition in these films.

Our experiments provide insight into the magnetic and electronic properties, and should guide detailed band structure calculations of the doped ETO/LTO system to give a clearer picture of the underlying mechanisms of these transitions and the formation of strongly-correlated system between polarized Eu 4$f$ and LHB. This may reveal additional emergent features between a Mott insulator and a band insulator, and broaden the application levels in similar oxide systems.

In conclusion, the evolution from AF band insulator ETO to Mott insulator LTO thin film is confirmed via transport measurements, with magnetic and metallic transitions from AFM to FM and PM. These phase transitions and transition temperatures depend sensitively on the La substitution level, while there is a slight discrepancy in magnetic states depending on the substituted elements and the form of crystals. The films in AFM and FM phases show clear AHE despite being induced by different mechanisms. Moreover, the drastic decrease of mobility near $x = 0.5$ could indicate an effective mass enhancement from strong on-site Coulomb interaction in the $t_{2g}$ band, or the interaction between Eu 4$f$ and the Ti LHB. This method of controlling the electronic and magnetic ground states may lead to the emergent functional devices and memories based on oxide systems [46–48].

**ACKNOWLEDGEMENTS**

This research was undertaken thanks in part to funding from the Max Planck-UBC-UTokyo Centre for Quantum Materials and the Canada First Research Excellence Fund, Quantum Materials and Future Technologies Program. The work at UBC was also supported by Natural Sciences and Engineering Research Council of Canada (NSERC) and Canada Foundation for Innovation (CFI). The part of experiments were performed at the Surface Science Facility of the REIXS beamline in the Canadian Light Source, which is funded by the Canada Foundation for Innovation, NSERC, the National Research Council of Canada, the Canadian Institutes of Health Research, the Government of Saskatchewan, Western Economic Diversification Canada, and the University of Saskatchewan. We thank Dr. Ryan Thorpe of the Rutgers Laboratory for Surface Modification (Department of Physics, Rutgers University, Piscataway NJ) for RBS measurements.

Corresponding Author

*E-mail: hshin@phas.ubc.ca

*E-mail: kzou@phas.ubc.ca

ORCID

Hyungki Shin: 0000-0001-8435-4479
Chong Liu: 0000-0002-2314-8687
Ronny Sutarto: 0000-0002-1969-3690
Bruce A. Davidson: 0000-0003-1616-9380
Ke Zou: 0000-0002-1181-1779